\documentclass{article}
\usepackage{PRIMEarxiv}

\usepackage{moreverb,url}
\usepackage{enumitem}
\usepackage[utf8]{inputenc} 
\usepackage[T1]{fontenc}    
\usepackage{url}            
\usepackage{booktabs}       
\usepackage{amsfonts}       
\usepackage{amsmath}
\usepackage{amssymb}
\usepackage{amsthm}
\usepackage{bbm}
\usepackage{bm}
\usepackage{color}
\usepackage{multicol}
\usepackage{multirow}
\usepackage{nicefrac}       
\usepackage{lipsum}
\usepackage{fancyhdr}       
\usepackage{graphicx}       
\usepackage[dvipsnames]{xcolor}
\usepackage{algorithmic}
\usepackage[ruled]{algorithm2e}

\usepackage[utf8]{inputenc}
\usepackage{caption}
\newcommand{\argmin}{\mathop{\rm argmin}\limits}

\newtheorem{theorem}{Theorem}[section]

\newtheorem{assumption}{Assumption}[section]
\graphicspath{{media/}}  
\newcommand{\redb}[1]{\textcolor{BrickRed}{\textbf{#1}}}
\newcommand{\indep}{\perp \!\!\! \perp}

\title{Quantile outcome adaptive lasso: covariate selection for inverse probability weighting estimator of quantile treatment effects
}

\author{
  Takehiro Shoji \\
  \texttt{Graduate School of Culture and Information Science, Doshisha University}\\
  \texttt{luckmt0107@gmail.com} \\
   \And
  Jun Tsuchida \\
  \texttt{Faculty of Data Science, Kyoto Women's University
}\\
  \AND
  Hiroshi Yadohisa \\
  \texttt{Faculty of Culture and Information Science, Doshisha University}\\
}

\begin{document}
\maketitle

\begin{abstract}
When using the propensity score method to estimate the treatment effects, it is important to select the covariates to be included in the propensity score model. The inclusion of covariates unrelated to the outcome in the propensity score model led to bias and large variance in the estimator of treatment effects. Many data-driven covariate selection methods have been proposed for selecting covariates related to outcomes. However, most of them assume an average treatment effect estimation and may not be designed to estimate quantile treatment effects (QTE), which is the effect of treatment on the quantiles of outcome distribution. 
In QTE estimation, we consider two relation types with the outcome as the expected value and quantile point. To achieve this, we propose a data-driven covariate selection method for propensity score models that allows for the selection of covariates related to the expected value and quantile of the outcome for QTE estimation. Assuming the quantile regression model as an outcome regression model, covariate selection was performed using a regularization method with the partial regression coefficients of the quantile regression model as weights. The proposed method was applied to artificial data and a dataset of mothers and children born in King County, Washington, to compare the performance of existing methods and QTE estimators. 
As a result, the proposed method performs well in the presence of covariates related to both the expected value and quantile of the outcome.
\end{abstract}


\section{Introduction}
The investigation of causal effects are a topic of interest in many research fields. 
To investigate the causal effects, the effect of treatment on individual outcomes was evaluated. 
Additionally, one may not only be interested in the average treatment effect on multiple targets but also in the effect of treatment on the upper or lower side of the distribution of outcome by evaluating quantile treatment effects (QTE).
Studies using QTE include an evaluation of the effect of maternal smoking during pregnancy on birth weight \cite{tang2021new} and the effect of job training on annual income \cite{frumento2012evaluating}.

When estimating QTE from the observed data, the distribution of covariates affecting the outcome may differ between the group receiving the treatment (treatment group) and group not receiving the treatment (control group). 
It is difficult to distinguish the effect of treatment from the effect of confounding variables when the distribution of confounding variables
differs between the two groups. In such cases, an adjustment (covariate adjustment) is necessary to ensure that the distribution of covariates affecting the outcome is equal among treatments. Several methods have been proposed to estimate quantile treatment effects from observed data with covariate adjustments. Using an estimator weighed by the propensity score \cite{firpo,zhang2012causal}

Selecting the covariates to be included in the propensity score model is an important issue when estimating the treatment effects using propensity scores. The covariates included in the propensity score model are well discussed \cite{brookhart2006variable,de2011covariate,greenland2008invited}. In fact, it has been reported that including covariates unrelated to both the outcome and treatment in the propensity score model can bias the estimator of the treatment effect and that including covariates related to the treatment reduces the efficiency of the estimator. Therefore, the inclusion of covariates in the propensity score model should be carefully considered. In many cases, selecting covariates to be included in a propensity model is difficult a priori. To address this issue, several data-driven covariate selection methods have been proposed in the context of average treatment effects using propensity scores \cite{vansteelandt2012model,koch2018covariate,shortreed2017outcome}.

Although covariate selection methods for estimating average treatment effects have been well discussed, covariate selection methods for QTE estimation have not been sufficiently discussed. QTE estimation considers two possible relationships for the outcome, expected value and quantile value. 
In a linear model, it is easy to imagine that when it is related to the expected value, it is also related to the quantile. 
However, some variables were only related to the quantiles. 
For example, in the case of normal distribution, the covariate associated with variance is associated with its conditional quantile. Thus, existing covariate selection methods for average treatment effects are unsuitable for selecting covariates related to the outcome quantile. 

We propose a covariate selection method called Quantile Outcome Adaptive Lasso (QOAL) for QTE estimation using the propensity score method. Specifically, we modified the weights of regularization terms in the Outcome Adaptive Lasso\cite{shortreed2017outcome} covariate selection method, which assumes an average treatment effect. In OAL, the weights of regularization terms are partial regression coefficients of the multiple regression model, with the outcome as the objective variable. In QOAL, the weights of the regularization terms are partial regression coefficients of the quantile regression model\cite{koenker1978regression}, with the outcome as the objective variable. Using the partial regression coefficients of quantile regression as weighting terms, the proposed method is expected to select covariates associated with outcome quantiles.

This study proposes a covariate selection method for quantile treatment effect estimation. Section 2 describes the details of QOAL, a covariate selection method for quantile treatment effect estimation; Section 3 reports the results of numerical experiments to evaluate the performance of QOAL; Section 4 reports the results of applying the proposed method to real data; and Section 5 concludes the paper.

\section{QOAL}
\subsection{Estimand}
To define QTE, let $Y(j)\,(j=0,1)$ be the potential outcome\cite{rosenbaum1983central} and assume that the relationship between the observed and potential outcome is $Y=ZY(1) +(1-Z) Y(0)$ using the binary treatment variable $Z\in\{0,\,1\}$.
QTE is the difference between $\tau \in(0,\,1)$\, the quartile point of distribution of the potential outcome for the treated and untreated groups, and it is defined by the following: equation\cite{doksum1974}:
\begin{align*}
 {\rm{QTE}}=q_{1,\,\tau}-q_{0,\,\tau},
\end{align*}
where $q_{j,\,\tau}$ is defined as 
\begin{align}
\label{quantiledef}
    q_{j,\tau}\equiv \underset{q}{\rm{inf}}\,\{{\rm{Pr}}[Y(j) \leq q] \geq\tau\} \quad j=0,\,1.
\end{align}

\subsection{Propensity score}
Let $n$ be the sample size, $Y$ be a random variable representing the observed outcome, and $y_{i}$ be the realization of the outcome for target $i$. Let $Z$ be a binary (0: no treatment, 1: treatment) random variable (treatment variable) representing the treatment and $z_{i}$ be the realization of the treatment variable for target $i$. Let $\bm{X}=\,(X_{1},\,X_{2},\,\ldots,\,X_{p})'\,\in \mathbb{R}^{p}$ be the covariate vector, where $\bm{X}'$ denotes the transposition of $\bm{X}$.
Let $\bm{x}_{i}=\,(x_{i1},\,x_{i2},\,\ldots,\,x_{ip})'$ be the realization of the covariate vector of target $i$. The pair of observed random variables ($Y_i,\, Z_i ,\,\bm{X}'_i$) is assumed independent and identically distributed.
To estimate QTE, we make two assumptions, assignment of a strongly ignorable treatment and uniqueness of the quantile.

\begin{assumption}Strong ignorability\cite{rosenbaum1983central}\\
\label{asm:1}
The following condition holds for all $\bm{X} \in \mathcal{X}$, where $\mathcal{X}$ denotes support provided by $\bm{X}$.
\begin{enumerate}
    \item [\rm{(i) }]$(Y(1),\,Y(0))$ and the treatment $Z$ are conditionally independent under the condition of the covariate $\boldsymbol{X}$.
    \begin{align*}
        (Y(1),\, Y(0)) \indep Z \mid \bm{X}.
    \end{align*}
    \item [\rm{(ii) }] Conditional on the covariate $\bm{X}=\bm{x}$, the probability of assignment to a treatment is greater than $0$ and less than $1$.
    \begin{align*}
        0<{\rm{Pr}}(Z=1 \mid \bm{X}=\bm{x} ) <1.
    \end{align*}
\end{enumerate}
\end{assumption}
\noindent
  Consider two quantiles, $q_{1,\,\tau}$ and $q_{0,\,\tau}$ defined by Equation (\ref{quantiledef}). We assume that these two quantiles are unique to any $\,\tau \in (0,\,1)$.

\begin{assumption}Uniqueness of quantiles\cite{firpo} \mbox{}\\
\label{asm:2}
For $j=0,\,1$, $Y(j) $ is a continuous-valued random variable with $\mathbb{R}$ as its support and the following holds: 

There exists a non-empty set $\mathcal{Y}_{1}, \,\mathcal{Y}_{0}$ such that $\mathcal{Y}_{j}=\{\tau \in (0,1) ;\, {\rm{Pr}}[Y(j) \leq q_{j,\tau} - c]<{\rm{Pr}}[Y(j) \leq q_{j,\tau} + c\,],\, \forall c \in \mathbb{R},\, c>0 \}$.
\end{assumption}

Next, we defined the propensity scores.
The probability that a subject is assigned to a treatment conditional on the covariate $\bm{X}=\bm{x}$ is called the propensity score and defined by the following equation:
\begin{align*}
 \pi(\bm{x}) ={\rm{Pr}}(Z=1 \mid \bm{X}=\bm{x}).
\end{align*}
The propensity scores are generally unknown. We used a logistic regression model between the treatment variables and covariates (propensity score model).
\begin{align}\nonumber\label{eq:logit}
    {\rm{logit}}\{\pi(\bm{x}) \}&={\rm{logit}}\{{\rm{Pr}}(Z=1 \mid \bm{X}=\bm{x}) \} \\
    &=\bm{x}'\bm{\alpha},
\end{align}
where ${\rm{logit}}(p) ={\rm{log}}\left(p/(1-p) \right)$ and $\rm{\alpha}=\,(\alpha_{1},\,\alpha_{2},\,\ldots,\alpha_{p})'$ is the partial regression coefficient vector and $\alpha_{p+1}$ is the intercept.
From Assumptions \ref{asm:1} and \ref{asm:2}, some methods have been proposed to estimate QTE \cite{firpo,zhang2012causal,callaway}. We estimate the quantile treatment effect using the method proposed by Firpo \cite{firpo}. This is because Firpo's method estimates QTE by weighted quantile regression, which can be easily applied to real data.
Firpo shows that under Assumptions 1 and 2, the quantile of the distribution of a potential outcome can be expressed as a function of the observable variable using propensity scores. See Firpo's study for further details \cite{firpo}.

This study proposes a covariate selection method for the propensity score model in Equation \eqref{eq:logit} assuming QTE estimation. 
Let $\mathcal{C}$ be an index set of confounding variables, $\mathcal{P}_{E}$ be an index set of covariates related to the expected value of the outcome, $\mathcal{P}_{Q}$ be an index set of covariates related to the quantile of the outcome, $\mathcal{F}$ be an index set of covariates related to the treatment, and $\mathcal{I}$ be an index set of covariates that are neither treatment nor outcome related.
We assume  that the number of covariates $p = |\mathcal{C}|+|\mathcal{P}_{E}|+|\mathcal{P}_{Q}|+|\mathcal{F}|+|\mathcal{I}|$, where $|\cdot|$ denotes the cardinality of a set.
In estimation of the average treatment effect, the bias is reduced when covariate $X_{\mathcal{C}}$ is included in the propensity score model, and the statistical efficiency is improved when $\mathcal{P}_{E}$ is included in the propensity score model.

The proposed covariate selection method estimated the following propensity scores:
\begin{align}\nonumber
    {\rm{logit}}\{\tilde{\pi}(\bm{x}_{i}) \}=\sum_{j\in \mathcal{C}}\tilde{\alpha}_{j}x_{ij}+\sum_{j\in \mathcal{P}_{E}\cup\mathcal{P}_{Q}}\tilde{\alpha}_{j}x_{ij},
\end{align}
where $\tilde{\pi}(\bm{x}_{i})$ denotes the propensity score model. Since
$\tilde{\pi}(\bm{x}_{i})$ differs from ${\pi}(\bm{x}_{i})$,
the bias in the estimator of propensity score based on $\tilde{\pi}(\bm{x}_{i})$ tends to be large when $|\mathcal{F}|$ is large or the distribution of covariates is skewed. This is because the true propensity score model estimates the nonzero regression coefficient as $\tilde{\alpha}_{j}=0 (j \in \mathcal{F})$.

\subsection{Model and Parameter Estimation of QOAL}

In this study, we assume that the $100\tau\%$ quantile of the outcome at the quantile level $\tau\,(0<\tau<1)$ can be expressed using covariates and treatment variables (outcome regression model) as follows:
\begin{align}\label{eq:quantile}
        Q_{\tau}(Y \mid \bm{X}=\bm{x},Z=z) =\bm{x}'\bm{\beta}+z\beta_{p+1}+\beta_{p+2} .
\end{align}
where $Q_{\tau}(Y \mid \bm{X}=\bm{x}, Z=z)$ is the conditional $100\tau\%$ quantile of $Y$ conditioning $\bm{X}=\bm{x}$,  $\bm{\beta}=\,(\beta_{1},\,\beta_{2},\,\ldots,\,\beta_{p})'$ is the partial regression coefficient vector corresponding to the covariates, $\beta_{p+1}$ is the intercept term, and $\beta_{p+2}$ is the partial coefficient corresponding to the treatment variable. The partial regression coefficients are estimated by minimizing the check function defined by $\rho_{\tau}(u) =u(\tau-\mathbbm{1}\{u \leq 0\})$.

QOAL uses the estimated regression coefficients of the outcome regression model in Equation \eqref{eq:quantile} to estimate the partial regression coefficient vector of the propensity score model.
\begin{align}\label{eq:qoal-cost}
    \widehat{\boldsymbol{\alpha}}=\underset{\boldsymbol{\alpha}}{\operatorname{argmin}}\left[\sum_{i=1}^{n}\left\{-z_{i}\left(\bm{x}_{i}' \boldsymbol{\alpha}\right) + 
    \log \left(1+e^{\bm{x}_{i}' \boldsymbol{\alpha}}\right) \right\} 
     +\lambda \sum_{j=1}^{p} \widehat{\omega}_{j}\left|\alpha_{j}\right| \right].
\end{align}
where $\lambda\geq0$ is the regularization parameter. $\widehat{\omega}_{j}$ is $\widehat{\omega}_{j}=|\hat{\beta}_{j}|^{-\eta}$ for any $\,\eta>1$, where $\hat{\beta}_{j}\,(j=1,2,\ldots,p)$ is the estimated partial regression coefficient of the $j$-th covariate in equation (\ref{eq:quantile}) and $\lambda\geq0$ is the regularization parameter. $\widehat{\omega}_{j}$ is $\widehat{\omega}_{j}=|\hat{\beta}_{j}|^{-\eta}$ using any $\,\eta>1$, and $\hat{\beta}_{j}\,(j=1,2,\ldots,p)$ is the estimated partial regression coefficient of the $j$th covariate in Eq. (\ref{eq:quantile}) for the $j$th covariate. We define $\eta_{r}$ to satisfy $\lambda_{r}n^{\eta_{r}/2-1}=n^{2}$.

The QOAL estimate $\hat{\alpha}_{j}(\rm{QOAL})$ satisfies the following properties:
\begin{theorem}[Consistency]\mbox{} \\
For $\eta >1$, $\lambda /\sqrt{n} \rightarrow 0$ and $\lambda n^{\eta/2 -1}\rightarrow\infty$. Under the regularity condition, the {\rm{Quantile outcome adaptive lasso}} estimate $\hat{\bm{\alpha}}({\rm{QOAL}})$ satisfies 
    \begin{align*}
    \lim_{n\rightarrow\infty}\Pr\left\{\hat{\alpha}_{j}({\rm{QOAL}})=0 \mid j \in \mathcal{A}^{c}\right\}=1
    \end{align*}
    Let ${\mathcal{A}^{c}}=\mathcal{F}\cup\mathcal{I}$ be the partial regression coefficient corresponding to the covariate that is unrelated to the outcome. Additionally, $\hat{\bm{\beta}}$ is a root $n$-consistent estimator. Refer to the Appendix for the regularity condition and proof.
\end{theorem}
The above theorem guarantees that as the sample size increases, QOAL estimates the partial regression coefficients that are not related to the quantile of the outcome to be zero. This does not guarantee that the partial regression coefficients associated with quantiles of the outcome are estimated as nonzero.

\subsection{Selection criteria for regularization parameters}
The candidate regularization parameter predefined by the analyst is $\Lambda=\{\lambda_{r} |r=1,2,\ldots,R\}$, where $\Lambda$ is the number of regularization parameters. The regularization parameter $\lambda_{r}$ , which minimizes the weighted average absolute sum (wAMD) among the candidates, was used in the analysis. $\lambda$ selected using wAMD minimizes the absolute difference between two groups of weighted covariates, where wAMD is defined as
\begin{align}\label{eq:wAMD}
        \mbox{wAMD}(\lambda_{r}) =\sum_{j=1}^{p}|\hat{\beta}_{j}|\Biggm|\frac{\sum_{i=1}^n\hat{\phi}_{i}^{(\lambda_{r}) }x_{ij}z_{i}}{\sum_{i=1}^n\hat{\phi}_{i}^{(\lambda_{r}) }z_{i}} 
        -\frac{\sum_{i=1}^n\hat{\phi}_{i}^{(\lambda_{r}) }x_{ij}(1-z_{i}) }{\sum_{i=1}^n\hat{\phi}_{i}^{(\lambda_{r}) }(1-z_{i}) }\Biggm|,
\end{align}

Note that $\hat{\phi}^{(\lambda_{r}) }_{i}$ is defined using the propensity score-fitted value $\hat{\phi}_{i}^{(\lambda_{r}) }$ when estimated with $\lambda_{r}$ as follows:
\begin{align*}
    \hat{\phi}_{i}^{(\lambda_{r}) }=\frac{z_{i}}{\hat{\pi}_{i}^{(\lambda_{r}) }(\bm{x}_{i}) }+\frac{1-z_{i}}{1-\hat{\pi}_{i}^{(\lambda_{r}) }(\bm{x}_{i}) }.
\end{align*}

\subsection{Algorithm of QOAL}
We described the procedure for estimating the parameters of QOAL. First, we estimated the partial regression coefficients of the outcome regression model. Next, we calculated partial regression coefficients of the propensity score model and estimated the propensity score using a penalized logistic regression with partial regression coefficients of the outcome regression model. The estimated propensity score was used to calculate the quantile treatment effect using Firpo's method.
$\hat{\bm{\alpha}}$ calculated using $\lambda_{r}$ that minimizes wAMD is the estimated partial regression coefficient of the propensity score model (\textbf{Algorithm 1}).

\begin{algorithm}[htb]
 \begin{algorithmic}
 \renewcommand{\algorithmicrequire}{\textbf{Input:}}
 \renewcommand{\algorithmicensure}{\textbf{Output:}}
 \REQUIRE $\bm{x}_{i},y_{i},z_{i},\,(i=1,2,\ldots,n) ,\Lambda $
 \ENSURE  $\hat{\bm{\alpha}}$
 \STATE\label{for}
 \vspace{+0.5mm}
 $(\hat{\bm{\beta}},\hat{\beta}_{p+1},\hat{\beta}_{p+2}) =\argmin_{\bm{\beta},\,\beta_{p+1},\,\beta_{p+2}}\sum_{i=1}^{n}\rho_{\tau}(y_{i}-\{{\bm{x}_{i}}'\bm{\beta}+\beta_{p+1} z_{i}+\beta_{p+2}\}). $\\
 \STATE 
 We initialize $\bm{\alpha}_{0} \in \mathbb{R}^{p}$ as $r=0$. \\
\STATE 
Set $\eta_{r}$ to satisfy $\lambda_{r}n^{\eta_{r}/2-1}=n^{2}$. \\
\FOR{$r = 1,\,2,\ldots,\,R$}
\FOR{$j=1,2,\ldots,p$}
\STATE
set $\widehat{\omega}_{j}=|\hat{\beta}_{j}|^{-\eta_{r}}$. \\
\ENDFOR
\STATE
 Calculate the partial regression coefficient $\bm{\alpha}_{r}$ using $\lambda_{r}$ from Equation (\ref{eq:qoal-cost}).  \\
\ENDFOR
\STATE 
We set $\hat{\bm{\alpha}}$ as $\bm{\alpha}_{r}$ corresponding to $\lambda_{r}$ minimizing Equation \eqref {eq:wAMD}. \\
\end{algorithmic} 
\caption{Algorithm of QOAL}\label{alg}
\end{algorithm}

\section{Simulation}
Numerical simulations are conducted to evaluate the performance of the proposed method. We generated a dataset for use in numerical experiments with reference to the OAL article \cite{shortreed2017outcome}. 
The 20 covariates of each target $i\,(=1,2,\ldots,n)$ are independent and identical standard normal distributions.
We denote the covariate vector by $\bm{X}_{i}=\,(X_{i1},X_{i2},\ldots,X_{i20})'$. 
The binary treatment variable $Z_{i}$ is generated from the Bernoulli distribution of $\mbox{logit}\{{\Pr}(Z_{i}=1\mid \bm{X}=\bm{X}_{i})\}=\sum_{j=1}^{20}\alpha_{j}{X}_{ij}$. $\alpha_{j}$ is the true value of the $j\,th(=1,2,\ldots,20)$ partial regression coefficient in the propensity score model. A continuous-valued outcome $Y$ is generated from the following four scenarios.

\begin{enumerate}[leftmargin=1.5cm]
    \item[\textbf{senario1}]Simple regression model
    \begin{align*}
        Y_{i}=2Z_{i}+\sum_{j=1}^{20}\beta_{j}X_{ij}+\varepsilon_{i}.
    \end{align*}
    \item[\textbf{senario2}]Models with heterogeneous error variance
    \begin{align*}
        Y_{i}=2Z_{i}+\sum_{j=1}^{20}\beta_{j}X_{ij}+(1+0.75(X_{i1}+X_{i10})) \varepsilon_{i}.
    \end{align*}
    \item[\textbf{senario3}]Model with interaction between treatment variables and covariates
    \begin{align*}
        Y_{i}=2Z_{i}(1+X_{i2})+\sum_{j=1}^{20}\beta_{j}X_{ij}+(1+0.75(X_{i1}+X_{i10})) \varepsilon_{i}.
    \end{align*}
    \item[\textbf{senario4}]Outcome regression models that differed by treatment
    \begin{align*}
         Y_{i}(1)=\sum_{j=1}^{20}\beta^{(1)}_{j}X_{ij}+(1+0.75(X_{i1}+X_{i10})) \varepsilon_{i}, \\
         Y_{i}(0)=\sum_{j=1}^{20}\beta^{(0)}_{j}X_{ij}+(1+0.75(X_{i1}+X_{i10})) \varepsilon_{i}.
    \end{align*}
\end{enumerate}
where $\varepsilon_{i}\overset{{\rm{i.i.d.}}}{\sim} N(0,1)$. The true values of partial regression coefficients for the outcome regression and propensity score models were as follows:
\begin{align*}
\bm{\alpha}&=(1,0.4,0.4,0,0,0,1,1.8,1.8,\overbrace{0,\ldots,0}^{11})',\quad 
\bm{\beta}=(0.6,0.6,0.2,0.6,0.6,0.6,\overbrace{0,\ldots,0}^{14})' \\
\bm{\beta}^{(1)}&=(0.6,0.6,0.2,0.6,0.6,0.6,\overbrace{0,\ldots,0}^{14})' ,\quad 
\bm{\beta}^{(0)}=(-0.6,0.6,0.2,0.6,0.6,-0.6,\overbrace{0,\ldots,0}^{14})' 
\end{align*}

Let $X_{1},\, X_{2},\, X_{3}$ be the confounding variables; $X_{4},\, X_{5},\, X_{6}$ be covariates related only to the outcome; $X_{7},\, X_{8},\, X_{9}$ be covariates related only to the treatment variable; and the remaining are irrelevant covariates related to neither the treatment variable nor the outcome. In scenarios 2, 3, and 4, $X_{10}$ is a covariate that does not affect the expected value of the outcome but does affect the variance of the outcome.

To determine the effect of sample size on performance, $n=500,1000$ were used. The number of iterations in the numerical experiments was set to 10000. The treatment effect estimators rBias, rRMSE, and SD were used as evaluation indices. rBias and rRMSE were defined as follows: 
\begin{align*}
    \mbox{rBias }&=\frac{1}{K}\sum_{k=1}^{K}\frac{t-f_{k}}{t}, \\
    \mbox{rRMSE }&=\sqrt{\frac{1}{K}\sum_{k=1}^{K}\left(\frac{t-f_{k}}{t}\right) ^2 },
\end{align*}
where $f_{k}$ is the estimated quantile treatment effect for the $100\tau \%$ quantile repeated $k\,(1\leq k \leq K)$ times and $t$ is the true value of the quantile treatment effect for the $100\tau \%$ quantile. The true value $t$ of the quantile treatment effect at the $100\tau \%$ quantile is the difference between the $100\tau\%$ points of the potential outcome obtained by generating 200,000 data samples from the true-outcome regression model.

Eight methods were used for comparison. The four methods with covariate selection are {\textbf{Lasso}} \cite{tibshirani1996regression}, Adaptive Lasso (\textbf{Adl}) \cite{zou2006adaptive}, Average treatment effect estimation assumed Outcome adaptive lasso ({\textbf{OAL}}) \cite{shortreed2017outcome}, and the proposed Quantile Outcome Adaptive Lasso (\textbf{QOAL}). {\textbf{Lasso}} and \textbf{Adl} are expected to select the covariates associated with treatment. The {\textbf{OAL} selects the covariate associated with the expected outcome value. For further details on {\textbf{Lasso}}, {\textbf{OAL}}, and {\textbf{Adl}}, please refer to their respective studies.

The four methods without covariate selection fixed the covariates to be included in the propensity score model. The methods that include covariates and confounding related only to the expected value of the outcome in the propensity score model (\textbf{Targ\_Ex}) include only confounding (\textbf{Conf}) and covariates related to the expected value of the outcome and treatment (\textbf{Pot\_Conf}). 

The results of Scenario 1 are listed in Table \ref{tab:sin1}. In Scenario 1, the outcome was generated using a normal regression model. Thus, Scenario 1 is ideal for \textbf{OAL}. 
The method with covariate selection produced smaller values for all evaluation indices of \textbf{OAL} at almost all quantiles. 
For the method without covariate selection, the values of rBias and rRMSE for \textbf{Targ\_Ex} are small. 
The covariate selection rates of \textbf{QOAL} and \textbf{OAL} in Scenario 1 are shown in Figure \ref{fig:sin1}. There were no differences in the covariate selection rates between the two groups.

The results for Scenario 2 are presented in Table \ref{tab:sin2}. Scenario 2 is the ideal situation for \textbf{QOAL} because the variance of the outcome depends on the covariates. 
\textbf{QOAL} has smaller SD and rRMSE values for all quantiles among the method with covariate selection. 
For some quantiles, \textbf{OAL} has smaller values of rbias in the $n=1000$ case. 
The performance of the method without variable selection is better for \textbf{Targ\_Ex}.
The covariate selection rates for \textbf{QOAL} and \textbf{OAL} in Scenario 2 are shown in Figure \ref{fig:sin2}. \textbf{OAL} and \textbf{QOAL} had high selection rates for confounding variables. The main difference between \textbf{QOAL} and \textbf{OAL} is the selection rate of the tenth covariate. This covariate does not affect the expected value of the outcome, but it does affect its variance. We observed that the tenth covariate was selected at a higher rate in the $25\%$ and $75\%$ percentile quartiles for \textbf{QOAL}. Therefore, the rRMSE and SD of \textbf{QOAL} were smaller than those of \textbf{QOAL}.

The results for Scenario 3 are listed in Table \ref{tab:sin3}. In Scenario 3, the outcome was generated using a model with an interaction between the treatment variable and covariates. 
For $n=500$, all the evaluated indices in \textbf{QOAL} are small among the methods with covariate selection.
The values of rBias for \textbf{Targ\_Ex} indices were smaller than those of the other methods without covariate selection under some conditions.

For $n=1000$, \textbf{OAL} had the best results for all evaluation indices among the methods with covariate selection in the $25\%$ quartile. On the other hand, \textbf{QOAL} had the best results for all evaluation indexes among the methods with covariate selection at the $50\%$ and $75\%$ percentile quartiles. The trend for the method without covariate selection does not differ from $n=500$.
The covariate selection rates for \textbf{QOAL} and \textbf{OAL} covariates in Scenario 3 are shown in Figure \ref{fig:sin3}. 
From Figure \ref{fig:sin3}, \textbf{QOAL} and \textbf{OAL} have a larger selection proportion of covariates that are irrelevant to the outcome than  Scenarios 1 and 2.

The results of Scenario 4 are listed in Table \ref{tab:sin4}. In Scenario 4, the outcome model differed depending on the treatment. 
Among the methods with covariate selection, \textbf{OAL} had a smaller SD and rRMSE at $25\%$ and $50\%$. On the other hand, \textbf{QOAL} has a smaller rRMSE of $75\%$ among the methods with covariate selection. 
The performance of \textbf{Targ\_Ex} was better than that of the methods without covariate selection. 
The covariate selection rates of \textbf{QOAL} and \textbf{OAL} are shown in Figure \ref{fig:sin4}. 
Both \textbf{QOAL} and \textbf{OAL} have low covariate selection rates according to the first and third covariates, whose signs of the partial regression coefficients of the outcome regression model differ between the treatment and control groups. 
The selection rate of the 10th covariate, which is related to the outcome quantile, was high for \textbf{QOAL}.

\clearpage
\begin{table}[tb]
\renewcommand{\arraystretch}{1.5}
\centering
\caption{Results of Scenario 1\\
    Scenario 1 generated the outcome from a normal regression model.}
\label{tab:sin1}
\footnotesize
\begin{tabular}{rrrrrrrrrrrrr}
\hline
&\multicolumn{1}{c}{}        & \multicolumn{3}{c}{$\tau=0.25$} &  & \multicolumn{3}{c}{$\tau=0.5$} &  & \multicolumn{3}{c}{$\tau=0.75$} \\ \cline{3-5} \cline{7-9} \cline{11-13} 
                           & & {rBias}       & {rRMSE}      & {SD}      &  & {rBias}      & {rRMSE}      & {SD}      &  & {rBias}       & {rRMSE}      & {SD}      \\ \hline
\multirow{8}{*}{{$n=500$}}  & \textbf{QOAL} & 0.20 & 1.40 & 1.98 &  & 0.61 & 1.23 & 1.73 &  & 0.28 & 1.42 & 2.02 \\
 & \textbf{OAL} & \redb{0.09} & \redb{1.33} & \redb{1.88} &  & \redb{0.14} & \redb{1.18} & \redb{1.67} &  & \redb{$-$0.03} & \redb{1.32} & \redb{1.87} \\
 & \textbf{Lasso} & 10.95 & 2.71 & 3.15 &  & 11.06 & 2.63 & 3.00 &  & 10.82 & 2.99 & 3.63 \\
 & \textbf{Ald} & 8.49 & 3.36 & 4.45 &  & 7.43 & 3.21 & 4.29 &  & 7.94 & 3.46 & 4.64 \\ \cline{2-13}
 & \textbf{Targ\_Var} & $-$0.33 & \redb{1.33} & \redb{1.88} &  & $-$0.22 & \redb{1.15} & \redb{1.63} &  & $-$0.54 & 1.31 & 1.85 \\
 & \textbf{Targ\_Ex} & $-$0.32 & \redb{1.33} & \redb{1.88} &  & $-$0.22 & \redb{1.15} & \redb{1.63} &  & \redb{$-$0.52} & \redb{1.30} & \redb{1.84} \\
 & \textbf{Conf} & \redb{$-$0.24} & 1.52 & 2.15 &  & \redb{$-$0.21} & 1.38 & 1.95 &  & $-$0.54 & 1.46 & 2.06 \\
 & \textbf{Pot\_Conf} & 1.30 & 3.43 & 4.84 &  & 0.62 & 3.64 & 5.15 &  & 1.66 & 3.81 & 5.39 \\ \hline\hline	
\multirow{8}{*}{{$n=1000$}} & \textbf{QOAL} & 0.52 & 1.00 & 1.41 &  & 0.42 & 0.84 & 1.19 &  & \redb{0.02} & 1.01 & 1.43 \\
 & \textbf{OAL} & \redb{0.32} & \redb{0.91} & \redb{1.29} &  & \redb{0.20} & \redb{0.82} & \redb{1.16} &  & $-$0.18 & \redb{0.97} & \redb{1.38} \\
 & \textbf{Lasso} & 7.93 & 2.39 & 2.98 &  & 7.68 & 2.34 & 2.93 &  & 7.38 & 2.47 & 3.17 \\
 & \textbf{Ald} & 6.87 & 2.51 & 3.28 &  & 7.07 & 2.30 & 2.93 &  & 6.76 & 2.49 & 3.26 \\ \cline{2-13}
 & \textbf{Targ\_Var} & \redb{0.08} & 0.90 & 1.27 &  & \redb{$-$0.02} & \redb{0.80} & \redb{1.14} &  & $-$0.36 & 0.96 & 1.36 \\
 & \textbf{Targ\_Ex} & \redb{0.08} & \redb{0.89} & \redb{1.26} &  & \redb{$-$0.02} & \redb{0.80} & \redb{1.14} &  & $-$0.36 & \redb{0.95} & \redb{1.35} \\
 & \textbf{Conf} & 0.21 & 1.02 & 1.44 &  & 0.05 & 0.95 & 1.35 &  & $-$0.27 & 1.05 & 1.49 \\
 & \textbf{Pot\_Conf} & 0.86 & 3.03 & 4.28 &  & 0.22 & 3.01 & 4.26 &  & \redb{0.13} & 3.29 & 4.66 \\ \hline
\end{tabular}
\end{table}

\begin{figure}[bt]
\centering
\includegraphics[width=1\hsize]{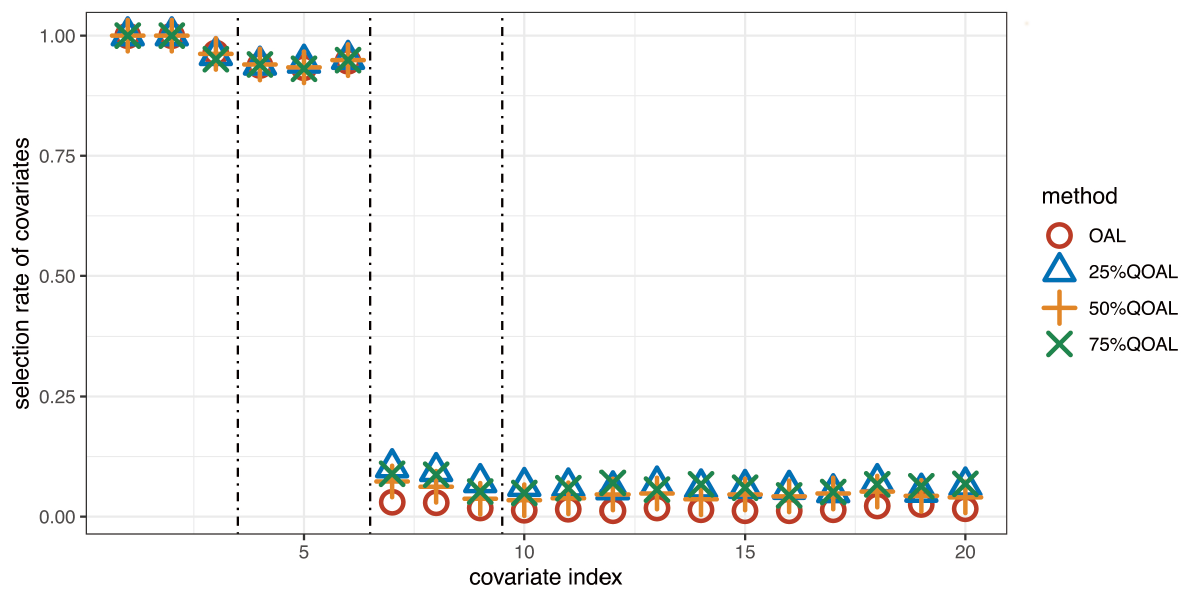}
\caption{Rate of covariates selection for $n=1000$ in scenario 1}
\label{fig:sin1}
\end{figure}

\begin{table}[tb]
\renewcommand{\arraystretch}{1.5}
\centering 
\caption{Results of Scenario 2 \\
Scenario 2 generated an outcome from a model with heterogeneous error variance.}
\label{tab:sin2}
\footnotesize
\begin{tabular}{rrrrrrrrrrrrr}
\hline
&\multicolumn{1}{c}{}        & \multicolumn{3}{c}{$\tau=0.25$} &  & \multicolumn{3}{c}{$\tau=0.5$} &  & \multicolumn{3}{c}{$\tau=0.75$} \\ \cline{3-5} \cline{7-9} \cline{11-13} 
                          & & {rBias}       & {rRMSE}      & {SD}      &  & {rBias}      & {rRMSE}      & {SD}      &  & {rBias}       & {rRMSE}      & {SD}      \\ \hline
\multirow{8}{*}{{$n=500$}}& \textbf{QOAL} & \redb{0.81} & \redb{1.10} & 2.21 &  & \redb{0.23} & \redb{0.82} & \redb{1.64} &  & \redb{0.08} & \redb{1.20} & \redb{2.40} \\
 & \textbf{OAL} & 0.90 & \redb{1.10} & \redb{2.20} &  & 0.71 & 0.91 & 1.81 &  & 0.29 & 1.26 & 2.52 \\
 & \textbf{Lasso} & 9.52 & 1.90 & 3.30 &  & 11.12 & 1.90 & 3.09 &  & 12.08 & 2.55 & 4.50 \\
 & \textbf{Ald} & 7.56 & 2.51 & 4.79 &  & 7.90 & 2.40 & 4.54 &  & 8.18 & 3.05 & 5.89 \\ \cline{2-13}
 & \textbf{Targ\_Var} & 0.15 & \redb{1.04} & 2.09 &  & $-$0.14 & \redb{0.81} & \redb{1.62} &  & $-$0.47 & \redb{1.14} & \redb{2.28} \\
 & \textbf{Targ\_Ex} & \redb{0.13} & \redb{1.04} & \redb{2.08} &  & $-$0.14 & \redb{0.81} & \redb{1.62} &  & $-$0.47 & \redb{1.14} & \redb{2.28} \\
 & \textbf{Conf} & 0.24 & 1.15 & 2.30 &  & \redb{$-$0.10} & 0.97 & 1.94 &  & \redb{$-$0.42} & 1.21 & 2.43 \\
 & \textbf{Pot\_Conf} & 1.18 & 2.63 & 5.26 &  & 0.57 & 2.89 & 5.79 &  & 1.34 & 3.37 & 6.74 \\
 \hline\hline	
\multirow{8}{*}{{$n=1000$}} & \textbf{QOAL} & \redb{0.29} & \redb{0.74} & \redb{1.48} &  & $-$0.16 & \redb{0.62} & \redb{1.25} &  & $-$0.33 & \redb{0.80} & \redb{1.60} \\
 & \textbf{OAL} & 0.57 & \redb{0.74} & \redb{1.48} &  & \redb{0.07} & 0.66 & 1.33 &  & \redb{0.02} & 0.83 & 1.66 \\
 & \textbf{Lasso} & 6.95 & 1.76 & 3.24 &  & 7.02 & 1.79 & 3.30 &  & 8.16 & 2.07 & 3.81 \\
 & \textbf{Ald} & 6.67 & 1.80 & 3.35 &  & 6.22 & 1.60 & 2.96 &  & 7.07 & 1.94 & 3.63 \\ \cline{2-13}
 & \textbf{Targ\_Var} & 0.08 & \redb{0.71} & \redb{1.42} &  & $-$0.32 & \redb{0.63} & \redb{1.26} &  & $-$0.56 & \redb{0.79} & \redb{1.57} \\
 & \textbf{Targ\_Ex} & \redb{0.05} & \redb{0.71} & 1.43 &  & $-$0.32 & \redb{0.63} & \redb{1.26} &  & $-$0.57 & \redb{0.79} & 1.58 \\
 & \textbf{Conf} & 0.18 & 0.80 & 1.60 &  & \redb{$-$0.25} & 0.73 & 1.47 &  & \redb{$-$0.54} & 0.84 & 1.68 \\
 & \textbf{Pot\_Conf} & 1.35 & 2.19 & 4.37 &  & $-$0.19 & 2.31 & 4.63 &  & $-$0.02 & 2.81 & 5.62 \\ \hline
\end{tabular}
\end{table}

\begin{figure}[bt]
\centering
\includegraphics[width=1\hsize]{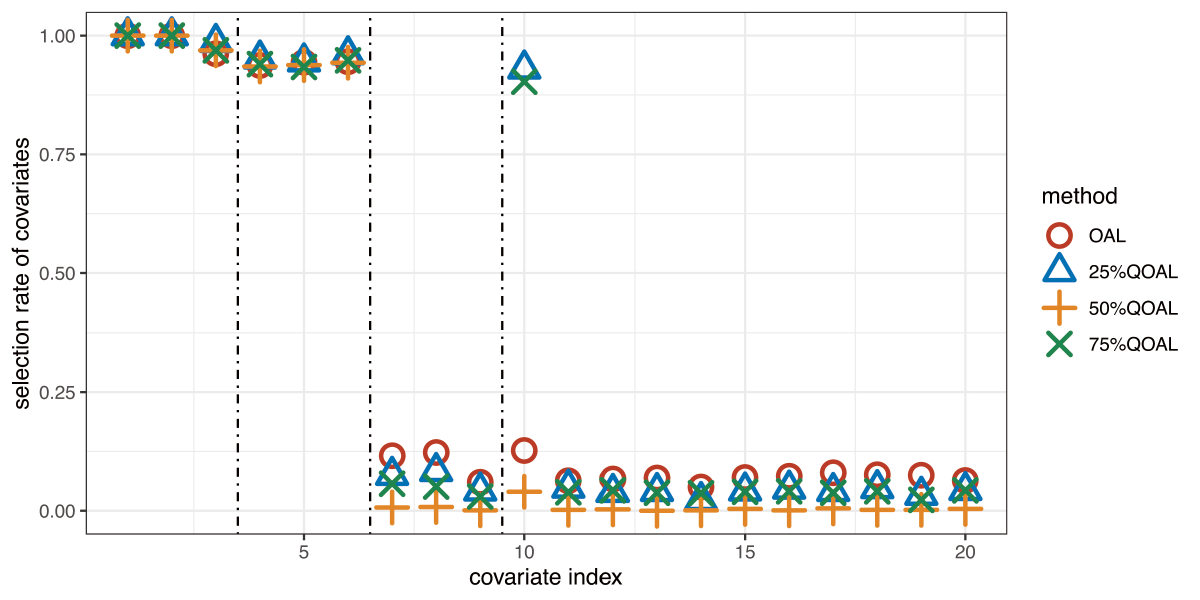}
\caption{Rate of covariates selection for $n=1000$ in scenario 2}
\label{fig:sin2}
\end{figure}

\begin{table}[tb]
\renewcommand{\arraystretch}{1.5}
\centering
\caption{Results of Scenario 3 \\
Scenario 3 was generated from a model with an interaction between the treatment variable and the covariates}
\label{tab:sin3}
\footnotesize
\begin{tabular}{rrrrrrrrrrrrr}
\hline
&\multicolumn{1}{c}{}        & \multicolumn{3}{c}{$\tau=0.25$} &  & \multicolumn{3}{c}{$\tau=0.5$} &  & \multicolumn{3}{c}{$\tau=0.75$} \\ \cline{3-5} \cline{7-9} \cline{11-13} 
                          & & {rBias}       & {rRMSE}      & {SD}      &  & {rBias}      & {rRMSE}      & {SD}      &  & {rBias}       & {rRMSE}      & {SD}      \\ \hline
\multirow{8}{*}{{$n=500$}} & \textbf{QOAL} & \redb{2.99} & \redb{2.92} & \redb{3.18} &  & \redb{0.93} & \redb{1.28} & \redb{2.45} &  & 1.20 & \redb{1.10} & \redb{3.09} \\
 & \textbf{OAL} & 3.22 & 2.93 & \redb{3.18} &  & 1.33 & 1.37 & 2.62 &  & \redb{1.03} & \redb{1.10} & 3.10 \\
 & \textbf{Lasso} & 34.99 & 5.66 & 4.87 &  & 18.97 & 3.03 & 4.54 &  & 13.90 & 2.42 & 5.60 \\
 & \textbf{Ald} & 22.75 & 7.25 & 7.54 &  & 11.44 & 3.47 & 6.30 &  & 8.34 & 2.67 & 7.16 \\ \cline{2-13}
 & \textbf{Targ\_Var} & \redb{0.03} & \redb{2.69} & \redb{2.95} &  & $-$0.36 & \redb{1.19} & \redb{2.28} &  & $-$0.03 & 0.98 & \redb{2.76} \\
 & \textbf{Targ\_Ex} & \redb{0.03} & \redb{2.69} & \redb{2.95} &  & $-$0.36 & \redb{1.19} & \redb{2.28} &  & $-$0.08 & \redb{0.97} & \redb{2.76} \\
 & \textbf{Conf} & 0.46 & 2.79 & 3.06 &  & \redb{$-$0.27} & 1.29 & 2.48 &  & \redb{0.01} & 1.01 & 2.87 \\ 
& \textbf{Pot\_Conf} & 3.59 & 8.01 & 8.76 &  & 0.72 & 4.06 & 7.81 &  & 1.68 & 3.02 & 8.53 \\
\hline\hline	
\multirow{8}{*}{{$n=1000$}} & \textbf{QOAL} & 2.30 & 1.97 & 2.15 &  & \redb{0.23} & \redb{0.94} & \redb{1.80} &  & \redb{0.13} & \redb{0.78} & \redb{2.19} \\
 & \textbf{OAL} & \redb{1.87} & \redb{1.96} & \redb{2.14} &  & 0.26 & 0.96 & 1.84 &  & 0.22 & 0.79 & 2.22 \\
 & \textbf{Lasso} & 25.71 & 5.34 & 5.13 &  & 12.78 & 2.55 & 4.25 &  & 9.03 & 1.89 & 4.70 \\
 & \textbf{Ald} & 18.27 & 5.12 & 5.23 &  & 9.20 & 2.39 & 4.25 &  & 6.37 & 1.76 & 4.63 \\ \cline{2-13}
 & \textbf{Targ\_Var} & 0.64 & \redb{1.78} & 1.95 &  & $-$0.61 & \redb{0.86} & \redb{1.64} &  & $-$0.39 & \redb{0.73} & \redb{2.05} \\
 & \textbf{Targ\_Ex} & \redb{0.61} & \redb{1.78} & \redb{1.94} &  & $-$0.61 & \redb{0.86} & \redb{1.64} &  & $-$0.39 & \redb{0.73} & 2.06 \\
 & \textbf{Conf} & 0.83 & 1.91 & 2.09 &  & \redb{$-$0.43} & 0.95 & 1.82 &  & \redb{$-$0.34} & 0.76 & 2.15 \\
 & \textbf{Pot\_Conf} & 5.89 & 6.22 & 6.78 &  & $-$0.55 & 3.41 & 6.56 &  & $-$0.25 & 2.44 & 6.90 \\ \hline																		
\end{tabular}
\end{table}

\begin{figure}[bt]
\centering
\includegraphics[width=1\hsize]{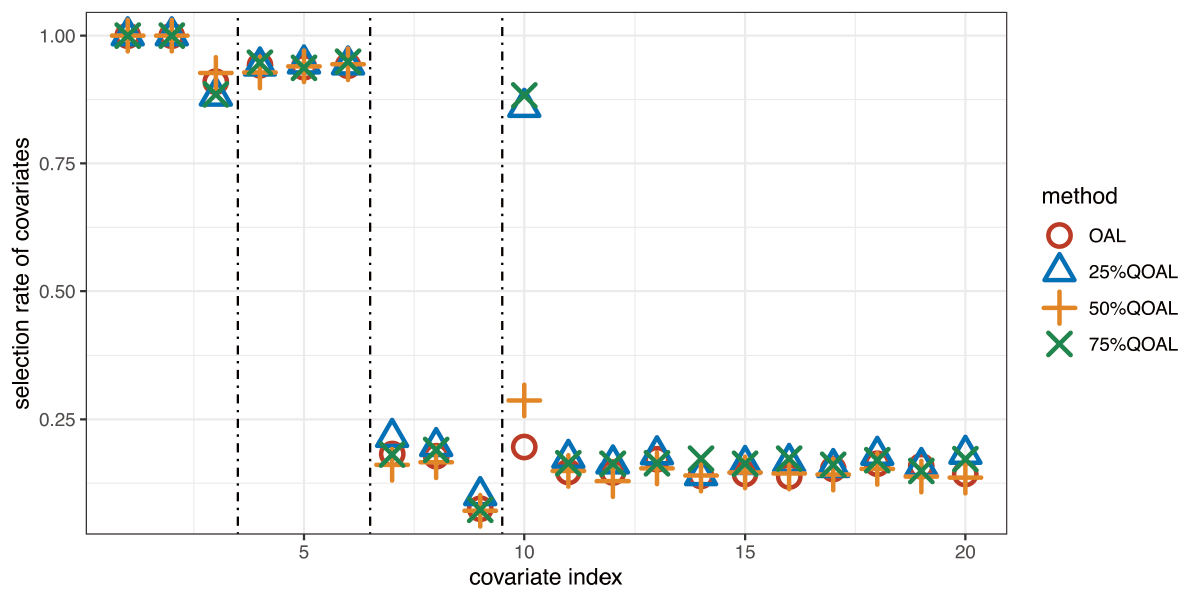}
\caption{Rate of covariates selection for $n=1000$ in scenario 3}
\label{fig:sin3}
\end{figure}

\begin{table}[tb]
\renewcommand{\arraystretch}{1.5}
\centering
\caption{Results of Scenario 4 \\ 
Scenario 4 generated outcomes from different models by treatment}
\label{tab:sin4}
\footnotesize
\begin{tabular}{rrrrrrrrrrrrr}
\hline
&\multicolumn{1}{c}{}        & \multicolumn{3}{c}{$\tau=0.25$} &  & \multicolumn{3}{c}{$\tau=0.5$} &  & \multicolumn{3}{c}{$\tau=0.75$} \\ \cline{3-5} \cline{7-9} \cline{11-13} 
                          & & {rBias}       & {rRMSE}      & {SD}      &  & {rBias}      & {rRMSE}      & {SD}      &  & {rBias}       & {rRMSE}      & {SD}      \\ \hline
\multirow{8}{*}{{$n=500$}}&  \textbf{QOAL} & 2.43 & 1.28 & 2.50 &  & 0.79 & \redb{1.05} & \redb{2.09} &  & \redb{$-$0.61} & 1.34 & 2.69 \\
 & \textbf{OAL} & \redb{1.59} & \redb{1.23} & \redb{2.44} &  & \redb{0.16} & 1.06 & 2.13 &  & $-$1.16 & \redb{1.30} & \redb{2.59} \\
 & \textbf{Lasso} & 5.67 & 1.97 & 3.78 &  & 5.17 & 1.76 & 3.38 &  & 6.03 & 2.06 & 3.95 \\
 & \textbf{Ald} & 6.94 & 2.68 & 5.17 &  & 5.71 & 2.26 & 4.38 &  & 5.42 & 2.55 & 4.99 \\ \cline{2-13}	
 & \textbf{Targ\_Var} & 0.19 & 1.12 & 2.24 &  & \redb{$-$0.54} & \redb{0.94} & \redb{1.89} &  & \redb{$-$0.70} & \redb{1.14} & \redb{2.27} \\
 & \textbf{Targ\_Ex} & \redb{0.09} & \redb{1.11} & \redb{2.21} &  & \redb{$-$0.54} & \redb{0.94} & \redb{1.89} &  & $-$0.77 & 1.15 & 2.30 \\
 & \textbf{Conf} & 0.10 & 1.21 & 2.42 &  & $-$0.58 & 1.03 & 2.07 &  & $-$0.90 & 1.23 & 2.47 \\
 & \textbf{Pot\_Conf} & 1.96 & 2.97 & 5.93 &  & $-$0.23 & 2.88 & 5.78 &  & $-$0.60 & 3.04 & 6.09 \\\hline\hline		
\multirow{8}{*}{{$n=1000$}} & \textbf{QOAL} & \redb{1.83} & 0.89 & 1.74 &  & 0.82 & 0.78 & 1.55 &  & \redb{$-$0.05} & \redb{0.90} & 1.81 \\
 & \textbf{OAL} & 1.86 & \redb{0.84} & \redb{1.63} &  & \redb{0.65} & \redb{0.75} & \redb{1.50} &  & $-$0.69 & \redb{0.90} & \redb{1.80} \\
 & \textbf{Lasso} & 4.70 & 1.72 & 3.30 &  & 4.81 & 1.63 & 3.13 &  & 4.72 & 2.06 & 4.02 \\
 & \textbf{Ald} & 5.91 & 1.93 & 3.67 &  & 5.69 & 1.65 & 3.11 &  & 5.38 & 1.79 & 3.42 \\ \cline{2-13}	
 & \textbf{Targ\_Var} & \redb{0.40} & \redb{0.80} & \redb{1.59} &  & $-$0.18 & \redb{0.69} & \redb{1.39} &  & $-$0.20 & \redb{0.81} & \redb{1.62} \\
 & \textbf{Targ\_Ex} & 0.42 & \redb{0.80} & \redb{1.59} &  & $-$0.18 & \redb{0.69} & \redb{1.39} &  & $-$0.24 & \redb{0.81} & \redb{1.62} \\
 & \textbf{Conf} & 0.55 & 0.85 & 1.70 &  & \redb{0.05} & 0.74 & 1.49 &  & \redb{$-$0.11} & 0.85 & 1.69 \\
 & \textbf{Pot\_Conf} & 1.51 & 2.52 & 5.02 &  & 1.28 & 2.31 & 4.64 &  & 0.61 & 2.63 & 5.27 \\ \hline
\end{tabular}
\end{table}

\begin{figure}[bt]
\centering
\includegraphics[width=1\hsize]{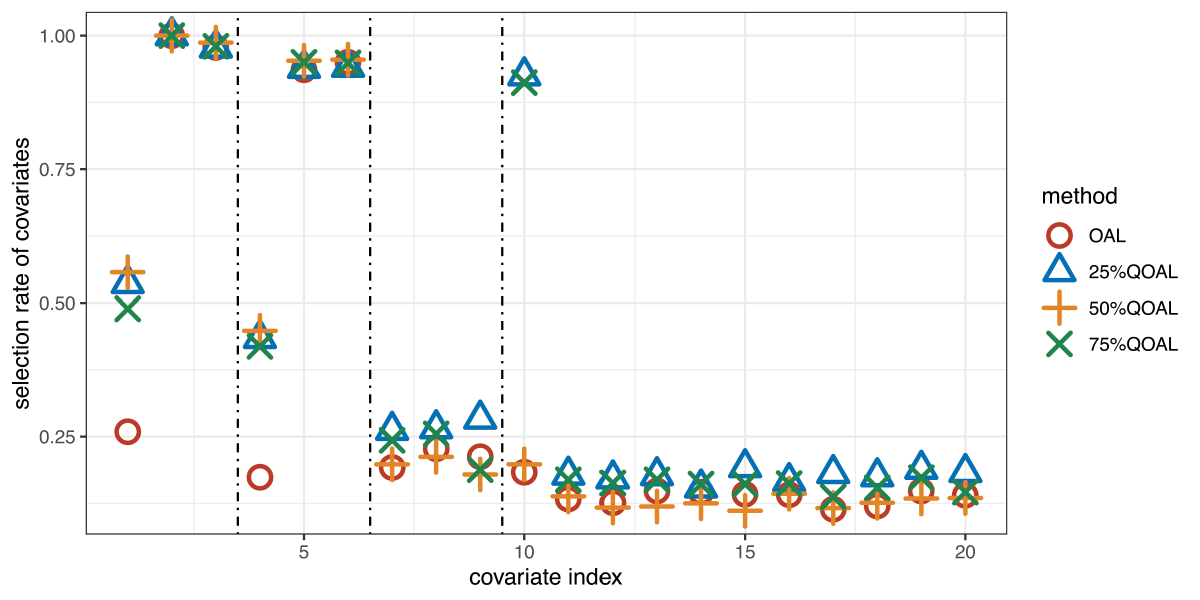}
\caption{Rate of covariates selection for $n=1000$ in scenario 4}
\label{fig:sin4}
\end{figure}

\clearpage
\section{Application}
This study investigated the effects of maternal smoking during pregnancy on birth weight using quantile treatment effects. Several studies have investigated the relationship between smoking during pregnancy and birth weight.
For example, prospective population studies have shown that smoking at conception leads to an average reduction in birth weight of $5\%$ \cite{brooke1989effects} and that smoking leads to a reduction in birth weight of approximately 250 - 640 g, depending on the maternal genotype \cite{wang2002maternal}.
Studies using quantile treatment effects \cite{xie2020multiply} suggest that maternal smoking during pregnancy leads to a weight loss of approximately 220 g in the $50\%$ quantile.

We evaluated QTE of smoking at conception on birth weight (g) using a dataset \cite{dukes2020obtain} of mothers and children born in King County, Washington, in 2001. The sample size was 2500. The dataset includes 15 covariates. The covariates included in the dataset are as follows: 
Table \ref{tab:1}. 
Among the covariates in table \ref{tab:1}, ``plural'' was excluded from the analysis because all observations are 1. ``smoker'' and ``drinker'' were excluded from the dataset because they are highly correlated with ``smokeN'' and ``drinkN.” The non-continuous values ``gender,” ``race,” ``married,” ``firstep,” and ``welfare'' were dummied and used in the analysis. The ``race'' category contains five races,``asian,” ``black,” ``hispanic,” ``other,” and ``white.”

\begin{table*}[htp]
  \caption{List of covariates used in Application}
  \label{tab:1}
  \renewcommand{\arraystretch}{1.2}
  \centering
  \small
  \begin{tabular}{ll}
    \hline 
``gender''   &     M = male, F = female baby \\
``plural''    &    1 = singleton, 2 = twin, 3 = triplet\\
``age''       &    mother's age in years\\
``race''      &    race categories (for mother)\\
``parity''    &    number of previous live born infants\\
``married''   &    Y = yes, N = no\\
``bwt''       &    birth weight in grams\\
``smokeN''    &    number of cigarettes smoked per day during pregnancy\\
``drinkN''    &    number of alcoholic drinks per week during pregnancy\\
``firstep''   &    1 = participant in program; 0 = did not participate\\
``welfare''   &    1 = participant in public assistance program; 0 = did not\\ 
``smoker''    &    Y = yes, N = no, U = unknown\\
``drinker''   &    Y = yes, N = no, U = unknown \\
``wpre''      &    mother's weight in pounds prior to pregnancy\\
``wgain''     &    mother's weight gain in pounds during pregnancy\\
``education'' &    highest grade completed (add 12 + 1 / year of college)\\
``gestation'' &    weeks from last menses to birth of child\\
    \hline
  \end{tabular}
\end{table*}

As we did not identify any relationship between the variables, we applied methods with covariate selection. We used \textbf{OAL} and \textbf{QOAL} as covariate selection methods because they perform well in the simulations. Sampling was repeated 100 times. The mean of estimated QTE by each method is shown in Table \ref{tab:2}. The results suggest that the $25\%$ quartile of QOAL leads to a weight loss of 136.86 g, $50\%$ quartile leads to a weight loss of 161.69 g, and $75\%$ quartiles of QOAL lead to a weight loss of 157.49 g. 

\begin{table*}[htb]
  \caption{the mean of estimated QTE of maternal smoking on birth weight}
  \renewcommand{\arraystretch}{1.5}
  \label{tab:2}
  \centering
    \begin{tabular}{llll} \hline
        & \multicolumn{1}{c}{$\tau=0.25$} &\multicolumn{1}{c}{$\tau=0.5$}&\multicolumn{1}{c}{$\tau=0.75$}  \\ \hline \hline
    {\textbf{QOAL}} & $-$120.25 & $-$171.31 & $-$163.35                    \\
    {\textbf{OAL}} & $-$110.81 & $-$173.21 & $-$158.42                        \\
    {\textbf{SimDiff}} &$-$227.81 & $-$194.96 & $-$212.95                      \\
    \hline
    \end{tabular}
\end{table*}

\section{Discussion}
This study proposes a covariate selection method for a propensity score model that assumes QTE estimation.
By using the partial regression coefficients of the model to estimate the conditional quantile of the outcome as weights for the regularization term, it was possible to select covariates related to the quantiles of the outcome.
The theoretical properties of QOAL guarantee that an increase in sample size does not select covariates that are irrelevant to either the treatment or the outcome. This does not guarantee selection associated with the outcome when the sample size increases.

Through simulation, we confirmed that QOAL performed better in QTE than the other methods, with covariate selection in many cases. 
Additionally, QOAL tended to select covariates related to the outcome quantile. 
The results of the simulation suggest that using QOAL tends to provide a better QTE estimation than using OAL when the relationship between the covariates and outcome is not obvious, although we examined the performance of these methods in a simple situation. 
However, if a linear model is assumed for the outcome regression model and the covariates associated with the expected value of the outcomes are known a priori, it is recommended to use a method without variable selection instead of a method with covariate selection.

We examined the effect of maternal smoking on birth weight using QTE with covariate selection. These results suggested that maternal smoking during pregnancy leads to weight loss in the upper quartiles. The results are similar to those of a previous study\cite{brooke1989effects}, in which maternal smoking during pregnancy was known to reduce birth weight by approximately $5\%$.

Four points should be noted when using the proposed method. 
First, the regularization method employed in QOAL estimates the regression coefficients of covariates related to the treatment to be 0. Thus, QOAL tends to misspecify the propensity score model. 
Particularly, if the proportion of covariates related only to treatment among all covariates is high, the difference between the estimated and true models of the propensity score will be large. This may have caused a bias in the treatment effect. 
Although the propensity score can be estimated for data with more covariates than the sample size, difference between the estimated and true models of propensity score may also be large. Hence, we carefully applied the proposed method to high-dimensional data and selected covariates for QTE estimation a priori.

Second, as in Scenario 3, when the outcome regression model has an interaction term between the treatment variable and covariates, the covariate selection rate related to the outcome is lower than that in the no-interaction case. Additionally, in Scenario 3, the selection rate of the covariate that was neither treatment nor outcome-related was higher. 
This may have caused a difference in the selection rates of the confounding variables. To achieve this, we extend the proposed method to consider the interaction term of the quantile regression.

Third, as in Scenario 4, we did not assume a situation in which the outcome regression models of the treatment and control groups differed. This may lead to instability in the selection of covariates and estimation of the treatment effect. 

Fourth, the estimation of the treatment effect in extreme quantiles may be unstable. 
This may be improved using the quantile treatment effect \cite{deuber2023estimation} for extreme quantile and quantile regression \cite{chernozhukov2005extremal} for the extreme quantile as weights for the regularization term. 
The performance of the proposed method was highly dependent on the quantile regression model, which is an outcome regression model.


\bibliographystyle{plain}  
\bibliography{references}

\section*{Appendix A: Proof}
We assumed the following regularity condition:
\begin{assumption}Regularity conditions\mbox{}  \label{as:regular}
    \begin{enumerate}
    \item The Fisher information matrix is positive definite.
    \[
        \mathbf{I}\left(\boldsymbol{\alpha}^*\right)=E\Big[\phi^{\prime \prime}\left(\bm{x}' \boldsymbol{\alpha}^*\right) \bm{x} \bm{x}'\Big],
    \]
    where $\phi(\theta) = \log(1 + \exp(\theta))$.
    \item $\phi(\bm{x}'\boldsymbol{\alpha}^*)$ is third-order partial differentiable with respect to $\bm{\alpha}$.
    \item For a neighborhood $\bm{\alpha}^{*}$ of $\bm{\alpha}$, there exists $M_{1}(\bm{x}),\,M_{2}(\bm{x})$ such that
    \[
    \left|\frac{\partial\phi(\bm{x};\bm{\alpha})}{\partial \alpha_{j}}\right|\leq M_{1}(\bm{x}),\quad
    \left|\frac{\partial^{2}\phi(\bm{x};\bm{\alpha})}{\partial \alpha_{j}\alpha_{\ell}}\right|\leq M_{1}(\bm{x}), \quad
    \left|\frac{\partial^{3}\phi(\bm{x};\bm{\alpha})}{\partial \alpha_{j}\alpha_{\ell}\alpha_{m}}\right|\leq M_{3}(\bm{x}),
    \]
    $\int M_{1}(\bm{x})d\bm{x} < \infty,$ and ${\rm{E}}[M_{2}(\bm{x}) | x_j,x_\ell,x_m]<\infty,\,(1\leq j,\ell,m\leq p)$.
    \end{enumerate}
\end{assumption}
We assume that $\hat{\bm{\beta}}$ is a root $n$-consistent estimator of $\hat{\bm{\beta}}^{*}$
Define $\bm{\alpha}=(\bm{\alpha}_{\mathcal{A}},\bm{\alpha}_{\mathcal{A}^{c}})=\bm{\alpha}^{*}+\bm{b}/\sqrt{n}$, where $\bm{\alpha}_{\mathcal{A}}$ is the partial regression coefficient corresponding to the covariate associated with the outcome and $\bm{\alpha}_{\mathcal{A}^{c}}$ is the partial regression coefficient corresponding to the covariate not associated with the outcome. The penalized log-likelihood function for the proof is defined as:
    \begin{align*}
        \Tilde{\ell}_{n}(\bm{\alpha})={\ell}_{n}(\bm{\alpha})-p_{n}(\bm{\alpha}).
    \end{align*}
    Let ${\ell}_{n}(\bm{\alpha})$ be the log-likelihood function of logistic loss.
    \begin{align*}
        p_{n}(\bm{\alpha})=\lambda_n \sum_{j=1}^p \hat{\omega}_j\left|\alpha_j\right|.
    \end{align*}

    Let $(\hat{\bm{\alpha}}_{\mathcal{A}},\bm{0})$ be the maximizer of penalized log-likelihood function $\tilde{\ell}_n\left(\boldsymbol{\alpha}_{\mathcal{A}}, \mathbf{0}\right)$.
    It suffices to show that in the neighborhood $\|\bm{\alpha}-\bm{\alpha}^{*}\|=O_{p}(n^{-1/2})$, $\tilde{\ell}_n\left(\boldsymbol{\alpha}_{\mathcal{A}}, \boldsymbol{\alpha}_{\mathcal{A}^{c}}\right) - \tilde{\ell}_n\left(\hat{\boldsymbol{\alpha}}_{\mathcal{A}}, \mathbf{0}\right)<0 $ with probability tending to 1 as $n\rightarrow\infty$.
\begin{align}
    \nonumber
    \tilde{\ell}_n\left(\boldsymbol{\alpha}_{\mathcal{A}}, \boldsymbol{\alpha}_{\mathcal{A}^{c}}\right) - \tilde{\ell}_n\left(\hat{\boldsymbol{\alpha}}_{\mathcal{A}}, \mathbf{0}\right) 
    &= \Big(\tilde{\ell}_n\left(\boldsymbol{\alpha}_{\mathcal{A}}, \boldsymbol{\alpha}_{\mathcal{A}^{c}}\right) - \tilde{\ell}_n\left(\boldsymbol{\alpha}_{\mathcal{A}}, \mathbf{0}\right)\Big) + \Big(\tilde{\ell}_n\left(\boldsymbol{\alpha}_{\mathcal{A}}, \boldsymbol{0}\right) - \tilde{\ell}_n\left(\hat{\boldsymbol{\alpha}}_{\mathcal{A}}, \mathbf{0}\right)\Big) \\
    \label{eq:consis_1}
    &\leq \Big(\tilde{\ell}_n\left(\boldsymbol{\alpha}_{\mathcal{A}}, \boldsymbol{\alpha}_{\mathcal{A}^{c}}\right) - \tilde{\ell}_n\left(\boldsymbol{\alpha}_{\mathcal{A}}, \mathbf{0}\right)\Big).
\end{align}
Consider the positivity and negativity of the right-hand side of the inequality \eqref{eq:consis_1}. 
By the definition of $\Tilde{\ell}_{n}(\cdot)$, it follows that
\begin{align*}
    \tilde{\ell}_n\left(\boldsymbol{\alpha}_{\mathcal{A}}, \boldsymbol{\alpha}_{\mathcal{A}^{c}}\right) - \tilde{\ell}_n\left(\boldsymbol{\alpha}_{\mathcal{A}}, \mathbf{0}\right) &= \Big(\ell_{n}\left(\boldsymbol{\alpha}_{\mathcal{A}}, \boldsymbol{\alpha}_{\mathcal{A}^{c}}\right) - \ell_{n}\left(\boldsymbol{\alpha}_{\mathcal{A}}, \mathbf{0}\right)\Big)  - \Big( p_n\left(\boldsymbol{\alpha}_{\mathcal{A}}, \boldsymbol{\alpha}_{\mathcal{A}^{c}}\right) - p_n\left(\boldsymbol{\alpha}_{\mathcal{A}}, \mathbf{0}\right)\Big).
\end{align*}
We first focus on the first term. As the log-likelihood function is differentiable, using the multivariate mean value theorem,
\begin{align}\label{eq:consis_2}\nonumber
    \ell_{n}\left(\boldsymbol{\alpha}_{\mathcal{A}}, \boldsymbol{\alpha}_{\mathcal{A}^{c}}\right) - \ell_{n}\left(\boldsymbol{\alpha}_{\mathcal{A}}, \mathbf{0}\right)&=\left[\frac{\partial \ell_n\left(\boldsymbol{\alpha}_{\mathcal{A}}, \boldsymbol{\xi}\right)}{\partial \boldsymbol{\alpha}_{\mathcal{A}}}\right]' (\boldsymbol{\alpha}_{\mathcal{A}}-\boldsymbol{\alpha}_\mathcal{A})+\left[\frac{\partial \ell_n\left(\boldsymbol{\alpha}_{\mathcal{A}}, \boldsymbol{\xi}\right)}{\partial \boldsymbol{\alpha}_{\mathcal{A}^{c}}}\right]' (\boldsymbol{\alpha}_{\mathcal{A}^{c}}-\bm{0}) \\
    &=\left[\frac{\partial \ell_n\left(\boldsymbol{\alpha}_{\mathcal{A}}, \boldsymbol{\xi}\right)}{\partial \boldsymbol{\alpha}_{\mathcal{A}^{c}}}\right]' \boldsymbol{\alpha}_{\mathcal{A}^{c}},
\end{align}
for certain $\|\bm{\xi}\|\leq\|\bm{\alpha}_{\mathcal{A}^{c}}\|$. Trigonometric inequalities were used.
\begin{align}\label{eq:consis_3}
    \left\|\frac{\partial \ell_n\left(\boldsymbol{\alpha}_{\mathcal{A}}, \boldsymbol{\xi}\right)}{\partial\boldsymbol{\alpha}_{\mathcal{A}^{c}}}-\frac{\partial \ell_n\left(\boldsymbol{\alpha}^{*}_{\mathcal{A}}, \boldsymbol{0}\right)}{\partial\boldsymbol{\alpha}_{\mathcal{A}^{c}}}\right\|\leq 
    \left\|\frac{\partial \ell_n\left(\boldsymbol{\alpha}_{\mathcal{A}}, \boldsymbol{\xi}\right)}{\partial\boldsymbol{\alpha}_{\mathcal{A}^{c}}}-\frac{\partial \ell_n\left(\boldsymbol{\alpha}_{\mathcal{A}}, \boldsymbol{0}\right)}{\partial\boldsymbol{\alpha}_{\mathcal{A}^{c}}}\right\|+\left\|\frac{\partial \ell_n\left(\boldsymbol{\alpha}_{\mathcal{A}}, \boldsymbol{0}\right)}{\partial\boldsymbol{\alpha}_{\mathcal{A}^{c}}}-\frac{\partial \ell_n\left(\boldsymbol{\alpha}^{*}_{\mathcal{A}}, \boldsymbol{0}\right)}{\partial\boldsymbol{\alpha}_{\mathcal{A}^{c}}}\right\| .
\end{align}
From the definition of $\ell(\cdot)$, we have
\begin{align*}
\ell_{n}(\bm{\alpha})=\sum_{i=1}^{n}\Big(z_{i}\bm{x}_{i}'\bm{\alpha}-\phi(\bm{x}_{i}'\bm{\alpha})\Big).
\end{align*}
The first term on the right-hand side of equation \eqref{eq:consis_3} is based on the mean value theorem, there exist $\bm{\xi}$ such that
\begin{align*}
    \left\|\frac{\partial \ell_n\left(\boldsymbol{\alpha}_{\mathcal{A}}, \boldsymbol{\xi}\right)}{\partial\boldsymbol{\alpha}_{\mathcal{A}^{c}}}-\frac{\partial \ell_n\left(\boldsymbol{\alpha}_{\mathcal{A}}, \boldsymbol{0}\right)}{\partial\boldsymbol{\alpha}_{\mathcal{A}^{c}}}\right\|&= \left\|\sum_{i=1}^n \Big(\frac{\partial \phi(\bm{x}_{i}'(\boldsymbol{\alpha}_{\mathcal{A}}, \boldsymbol{\xi}))}{\partial \bm{\alpha}_{\mathcal{A}^{c}}}-\frac{\partial \phi(\bm{x}_{i}'(\boldsymbol{\alpha}_{\mathcal{A}}, \boldsymbol{0}))}{\partial \bm{\alpha}_{\mathcal{A}^{c}}}\Big)\right\| \\
    &=\left\|\sum_{i=1}^{n}\frac{\partial^{2} \phi(\bm{x}_{i}'(\boldsymbol{\alpha}_{\mathcal{A}}, \boldsymbol{\eta}))}{\partial \bm{\alpha}^{2}_{\mathcal{A}^{c}}}\bm{\xi}\right\|.
\end{align*}
Based on the regularity conditions, we have
\begin{align*}
    \left\|\sum_{i=1}^{n}\frac{\partial^{2} \phi(\bm{x}_{i}'(\boldsymbol{\alpha}_{\mathcal{A}}, \boldsymbol{\eta}))}{\partial \bm{\alpha}^{2}_{\mathcal{A}^{c}}}\bm{\xi}\right\| \leq\Bigg[\sum_{i=1}^{n}M_{1}(\bm{x}_{i})\Bigg]\|\bm{\xi}\|.
\end{align*}
The second term on the right-hand side of \eqref{eq:consis_3} is similarly
\begin{align*}
    \left\|\frac{\partial \ell_n\left(\boldsymbol{\alpha}_{\mathcal{A}}, \boldsymbol{0}\right)}{\partial\boldsymbol{\alpha}_{\mathcal{A}^{c}}}-\frac{\partial \ell_n\left(\boldsymbol{\alpha}_{\mathcal{A}*}, \boldsymbol{0}\right)}{\partial\boldsymbol{\alpha}_{\mathcal{A}^{c}}}\right\|  \leq\Bigg[\sum_{i=1}^{n}M_{1}(\bm{x}_{i})\Bigg]\|\bm{\alpha}_{\mathcal{A}}-\bm{\alpha}^{*}_{\mathcal{A}}\|.
\end{align*}
From Equation \eqref{eq:consis_3}, we obtain 
\begin{align*}
    \left\|\frac{\partial \ell_n\left(\boldsymbol{\alpha}_{\mathcal{A}}, \boldsymbol{\xi}\right)}{\partial\boldsymbol{\alpha}_{\mathcal{A}^{c}}}-\frac{\partial \ell_n\left(\boldsymbol{\alpha}^{*}_{\mathcal{A}}, \boldsymbol{0}\right)}{\partial\boldsymbol{\alpha}_{\mathcal{A}^{c}}}\right\|&\leq 
    \left\|\frac{\partial \ell_n\left(\boldsymbol{\alpha}_{\mathcal{A}}, \boldsymbol{\xi}\right)}{\partial\boldsymbol{\alpha}_{\mathcal{A}^{c}}}-\frac{\partial \ell_n\left(\boldsymbol{\alpha}_{\mathcal{A}}, \boldsymbol{0}\right)}{\partial\boldsymbol{\alpha}_{\mathcal{A}^{c}}}\right\|+\left\|\frac{\partial \ell_n\left(\boldsymbol{\alpha}_{\mathcal{A}}, \boldsymbol{0}\right)}{\partial\boldsymbol{\alpha}_{\mathcal{A}^{c}}}-\frac{\partial \ell_n\left(\boldsymbol{\alpha}^{*}_{\mathcal{A}}, \boldsymbol{0}\right)}{\partial\boldsymbol{\alpha}_{\mathcal{A}^{c}}}\right\| \\
    &\leq\Bigg[\sum_{i=1}^{n}M_{1}(\bm{x}_{i})\Bigg]\|\bm{\xi}\|+\Bigg[\sum_{i=1}^{n}M_{1}(\bm{x}_{i})\Bigg]\|\bm{\alpha}_{\mathcal{A}}-\bm{\alpha}^{*}_{\mathcal{A}}\| \\
    &= \big\{\|\bm{\xi}\|+\|\bm{\alpha}_{\mathcal{A}}-\bm{\alpha}^{*}_{\mathcal{A}}\|\big\}O_{p}(n).
\end{align*}
For $j \in \mathcal{F}, \|\bm{\xi}\|\leq \|\bm{\alpha}_{\mathcal{F}}\|=O_{p}(n^{-1/2})$, we have Thus
\begin{align*}
     \left\|\frac{\partial \ell_n\left(\boldsymbol{\alpha}_{\mathcal{A}}, \boldsymbol{\xi}\right)}{\partial\boldsymbol{\alpha}_{\mathcal{A}^{c}}}-\frac{\partial \ell_n\left(\boldsymbol{\alpha}^{*}_{\mathcal{A}}, \boldsymbol{0}\right)}{\partial\boldsymbol{\alpha}_{\mathcal{A}^{c}}}\right\| \leq O_{p}(n^{-1/2}).
\end{align*}
For $j \in \mathcal{I}, \|\bm{\xi}\|\leq \|\bm{\alpha}_{\mathcal{I}}\|=O_{p}(1)$. Thus
\begin{align*}
     \left\|\frac{\partial \ell_n\left(\boldsymbol{\alpha}_{\mathcal{A}}, \boldsymbol{\xi}\right)}{\partial\boldsymbol{\alpha}_{\mathcal{A}^{c}}}-\frac{\partial \ell_n\left(\boldsymbol{\alpha}^{*}_{\mathcal{A}}, \boldsymbol{0}\right)}{\partial\boldsymbol{\alpha}_{\mathcal{A}^{c}}}\right\| \leq O_{p}(n).
\end{align*}
Therefore, for $j \in \mathcal{A}^{c}$, we have
\begin{align*}
     \left\|\frac{\partial \ell_n\left(\boldsymbol{\alpha}_{\mathcal{A}}, \boldsymbol{\xi}\right)}{\partial\boldsymbol{\alpha}_{\mathcal{A}^{c}}}-\frac{\partial \ell_n\left(\boldsymbol{\alpha}^{*}_{\mathcal{A}}, \boldsymbol{0}\right)}{\partial\boldsymbol{\alpha}_{\mathcal{A}^{c}}}\right\| \leq O_{p}(n).
\end{align*}
Applying these results to Equation \eqref{eq:consis_2}, we have
\begin{align*}
    \ell_{n}\left(\boldsymbol{\alpha}_{\mathcal{A}}, \boldsymbol{\alpha}_{\mathcal{A}^{c}}\right) - \ell_{n}\left(\boldsymbol{\alpha}_{\mathcal{A}}, \mathbf{0}\right)&=\left[\frac{\partial \ell_n\left(\boldsymbol{\alpha}_{\mathcal{A}}, \boldsymbol{\xi}\right)}{\partial \boldsymbol{\alpha}_{\mathcal{A}^{c}}}\right]' \boldsymbol{\alpha}_{\mathcal{A}^{c}} \\
    &=\sum_{j \in \mathcal{A}^{c}}-|\alpha_{j}|O_{p}(n).
\end{align*}
Since,
\begin{align*}
    p_{n}(\bm{\alpha}_{\mathcal{A}},\bm{\alpha}_{\mathcal{A}^{c}})-p_{n}(\bm{\alpha}_{\mathcal{A}},\bm{0})=\lambda_{n}\sum_{j \in \mathcal{A}^{c}}\hat{\omega}_j\left|\alpha_j\right|,
\end{align*}
we have 
\begin{align}\label{eq:consis_4}
    \tilde{\ell}_n\left(\boldsymbol{\alpha}_{\mathcal{A}}, \boldsymbol{\alpha}_{\mathcal{A}^{c}}\right) - \tilde{\ell}_n\left(\boldsymbol{\alpha}_{\mathcal{A}}, \mathbf{0}\right) &=\sum_{j \in \mathcal{A}^{c}}\Big(-|\alpha_{j}|O_{p}(n)-\lambda_{n}\hat{\omega}_j\left|\alpha_j\right|\Big).
\end{align}
The proof is complete because the probability that Equation \eqref{eq:consis_4} is negative when $n\rightarrow\infty$ is 1.
\end{document}